\documentclass[twocolumn,showpacs,superscriptaddress, final, numbers, sort&compress, pra]{revtex4-1}

\usepackage{color,amsmath,amssymb,amsthm,times,graphics,graphicx,bm,bbm,dcolumn}
\usepackage{epsfig}
\usepackage{graphicx}
\usepackage[colorlinks,urlcolor=black,citecolor=blue,linkcolor=black]{hyperref}
\setcitestyle{round}

\usepackage[resetlabels,labeled]{multibib}
\newcites{SM}{References}

\newcommand{\ket}[1]{\mbox{\ensuremath{\vert #1 \rangle}}}

\newcommand{\previous}[1]{}

\newcommand{\beq}{\begin{equation}}
\newcommand{\eeq}{\end{equation}}

\begin{document}

\title{Strongly correlated superfluid order parameters from dc Josephson supercurrents}

\author{W. J. Kwon}
\affiliation{Istituto Nazionale di Ottica del Consiglio Nazionale delle Ricerche (CNR-INO), 50019 Sesto Fiorentino, Italy}
\affiliation{European Laboratory for Nonlinear Spectroscopy (LENS), 50019 Sesto Fiorentino, Italy}
\author{G. Del Pace}
\affiliation{European Laboratory for Nonlinear Spectroscopy (LENS), 50019 Sesto Fiorentino, Italy}
\affiliation{Department of Physics and Astronomy, University of Florence, 50019 Sesto Fiorentino,Italy}
\author{R. Panza}
\affiliation{Istituto Nazionale di Ottica del Consiglio Nazionale delle Ricerche (CNR-INO), 50019 Sesto Fiorentino, Italy}
\affiliation{European Laboratory for Nonlinear Spectroscopy (LENS), 50019 Sesto Fiorentino, Italy}
\author{M. Inguscio}
\affiliation{Istituto Nazionale di Ottica del Consiglio Nazionale delle Ricerche (CNR-INO), 50019 Sesto Fiorentino, Italy}
\affiliation{European Laboratory for Nonlinear Spectroscopy (LENS), 50019 Sesto Fiorentino, Italy}
\affiliation{Department of Engineering, Campus Bio-Medico University of Rome, 00128 Rome, Italy}
\author{W. Zwerger}
\affiliation{Physics Department, Technische Universit\"at M\"unchen, 85747 Garching, Germany}
\author{M. Zaccanti} 
\affiliation{Istituto Nazionale di Ottica del Consiglio Nazionale delle Ricerche (CNR-INO), 50019 Sesto Fiorentino, Italy}
\affiliation{European Laboratory for Nonlinear Spectroscopy (LENS), 50019 Sesto Fiorentino, Italy}
\author{F. Scazza}
\affiliation{Istituto Nazionale di Ottica del Consiglio Nazionale delle Ricerche (CNR-INO), 50019 Sesto Fiorentino, Italy}
\affiliation{European Laboratory for Nonlinear Spectroscopy (LENS), 50019 Sesto Fiorentino, Italy}
\author{G. Roati}
\affiliation{Istituto Nazionale di Ottica del Consiglio Nazionale delle Ricerche (CNR-INO), 50019 Sesto Fiorentino, Italy}
\affiliation{European Laboratory for Nonlinear Spectroscopy (LENS), 50019 Sesto Fiorentino, Italy}

\begin{abstract}
The dc Josephson effect provides a powerful phase-sensitive tool for investigating superfluid order parameters. We report on the observation of dc Josephson supercurrents in strongly interacting fermionic superfluids across a tunnelling barrier in the absence of any applied potential difference. For sufficiently strong barriers, we observe a sinusoidal current-phase relation, in agreement with Josephson's seminal prediction. We map out the zero-resistance state and its breakdown as a function of junction parameters, extracting the Josephson critical current behaviour. By comparing our results with an analytic model, we determine the pair condensate fraction throughout the Bardeen-Cooper-Schrieffer -- Bose-Einstein condensation crossover. Our work suggests that coherent Josephson transport may be used to pin down superfluid order parameters in diverse atomic systems, even in the presence of strong correlations.
\end{abstract}

\maketitle

When two superfluids or superconductors are weakly coupled through an insulating potential barrier, a dissipationless current $I_s$ can flow from one to the other, sustained merely by the relative phase difference $\varphi$ between the two order parameters \cite{Josephson1962,Anderson1964}.
This phenomenon, known as the dc Josephson effect \cite{Josephson1962,Likharev1979}, represents a paradigmatic manifestation of the macroscopic quantum phase coherence of any condensed state, and it is at the basis of remarkable applications \cite{Jaklevic1964,BaroneBook,Sato2011,Datta1997}, e.g.~in the field of metrology for high-precision measurements \cite{Popel1992,Clarke2004}. 
A distinctive feature of any Josephson junction is the link between the supercurrent $I_s$ and $\varphi$, namely the current-phase relation $I_s(\varphi)$ and, associated with it, the existence of a maximum current $I_c$. 
For barriers with sufficiently low transmission as those originally considered by Josephson \cite{Josephson1962}, where the tunnelling process can be treated as a perturbation, a simple sinusoidal current-phase relation $I_s(\varphi) = I_c \sin(\varphi)$ holds \cite{Anderson1964,Bloch1970}. An externally imposed current $I_\mathrm{ext}$ can flow without establishing any potential drop across the junction only if $\vert I_\mathrm{ext} \vert \leq I_c$, where $I_c$ is the Josephson critical current. The most striking signature of $I_c$ is therefore contained in the current-potential characteristic \cite{Tinkham,Anderson1969}. Such $I$-$V$ curve is routinely measured in current-biased superconducting Josephson junctions (SJJs), where a zero-voltage branch (the so-called dc branch) at $|I_\mathrm{ext}| \leq I_c$ can be clearly distinguished from a resistive finite-voltage branch at $|I_\mathrm{ext}| > I_c$. 
Acting as a precise interferometric probe, Josephson supercurrents offer a unique tool to disclose the nature and the symmetry of superfluid or superconducting order parameters, e.g.~the $d$-wave pairing symmetry in cuprate superconductors \cite{VanHarlingen1995,Tsuei2000}, and may become fundamental in the quest for Majorana bound states hosted by topological superconducting wires \cite{Kitaev2001,Jiang2011}. 

Coherent supercurrents have been extensively investigated with atomic Bose-Einstein condensates \cite{Albiez2005,Levy2007,LeBlanc2011,Ramanathan2011,Ryu2013,Jendrzejewski2014,Eckel2014a,Eckel2014b,Spagnolli2017,Pigneur2018}, offering exciting perspectives for atomtronics \cite{Chien2015}. On the other hand, ultracold Fermi gases offer the unique possibility to explore superfluid transport from the Bardeen-Cooper-Schrieffer (BCS) limit of weakly bound fermion pairs to a Bose-Einstein condensate (BEC) of tightly bound molecules, crossing over the intermediate universal, strongly correlated unitary regime \cite{Zwerger2012}. The dynamics of weakly connected fermionic superfluids is however fundamentally influenced and complicated by the inherent strong inter-particle interactions \cite{Stadler2012,Husmann2015,Valtolina2015,Krinner2016,Lebrat2018}, necessary also to achieve pair condensation at experimentally accessible temperatures \cite{Zwerger2012}. %
While the connection between the Josephson current and the superfluid order parameter is theoretically well established in both BCS and BEC limits \cite{Ambegaokar,Meier2001}, for crossover superfluids $I_c$ has been numerically calculated only at the mean-field level \cite{Spuntarelli2007,Zou2014}, where such a connection is not explicit, and it has not been experimentally explored. A direct measure of the Josephson critical current in strongly correlated superfluids is highly relevant \emph{per se}, and it may grant access to the order-parameter amplitude -- i.e., the pair condensate density \cite{Yang1962,Astrakharchik2005,Zaccanti2019} -- whose experimental determination has been so far indirect \cite{Zwierlein2004,Regal2004,Horikoshi2010} and somewhat inconclusive. %

\begin{figure*}[htbp]
\centering
{\includegraphics[width=\textwidth]{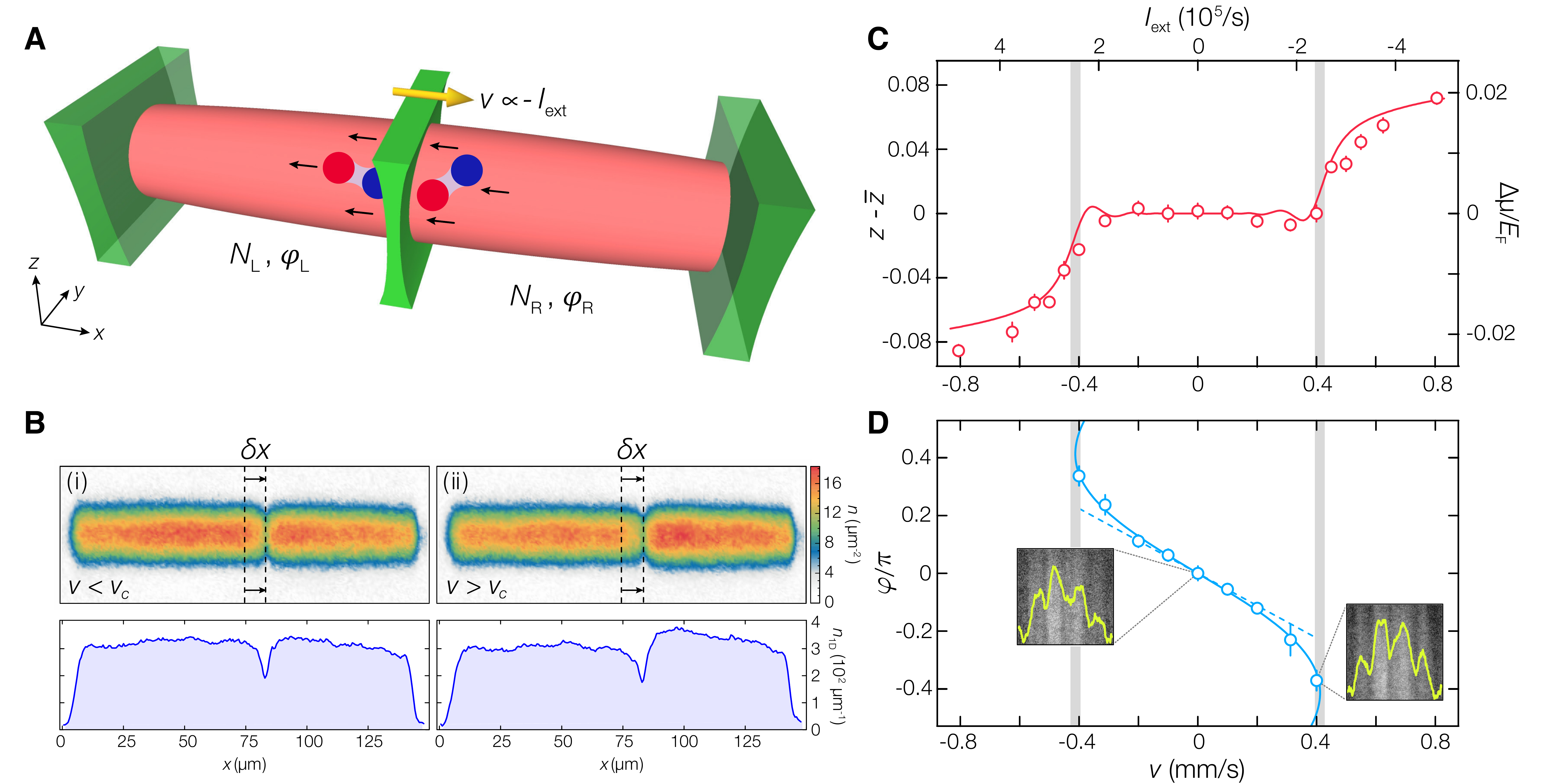}}
	\caption{Characterization of a current-biased atomic Josephson junction. (\textbf{A}) The Josephson junction is realized by weakly coupling two superfluid reservoirs (L, left; R, right) of $^6$Li fermion pairs through a thin DMD-generated, tunable repulsive barrier. An external current is imposed by translating the tunnelling barrier at constant velocity. (\textbf{B}) The dynamics is monitored by recording the number imbalance $z$ through \textit{in-situ} absorption imaging. (i) For currents below a critical value, pairs coherently tunnel through the barrier, maintaining a zero chemical-potential difference between the reservoirs. (ii) Conversely, above the critical current, the superfluid is compressed into the smaller reservoir. (\textbf{C}) Experimental $I$-$\Delta\mu$ characteristic of the junction for $(k_Fa)^{-1} \simeq 4.2$,  $w=0.95 \, \mu$m, and $V_0 \simeq 0.6E_F \simeq 1.8\,\mu$, where $\mu$ is the superfluid bulk chemical potential. The solid line denotes the RCSJ-model solution fit, while the shaded vertical lines represent the standard confidence interval of the extracted $I_c$. The small oscillatory behavior observed in the model just below $I_c$ stems from small-amplitude plasma oscillations excited by the non-adiabatic ramp-up of the applied current \cite{SM}. (\textbf{D}) Experimentally determined current-phase relation $I(\varphi)$ of the junction, recorded under the same conditions of panel (\textbf{C}) through matter-wave interference (see insets). The solid line is a sinusoidal fit to all data points including a small second-harmonic contribution $\sin{2\varphi}$, whereas the dashed line is a linear fit of the five central data points. Error bars in panels (\textbf{C-D}) indicate standard errors of the mean over $\sim10$ experimental realizations.}
\label{Fig1}
\end{figure*}

%
%
In this work, we observe the dc Josephson effect in a tunable, nearly ideal Josephson junction connecting two strongly correlated superfluids of ultracold fermionic atoms. 
By imparting a controlled current through strong and thin tunneling barriers, we map out the current-chemical potential $I$-$\Delta\mu$ relation throughout the BCS-BEC interaction crossover, which we find to closely parallel the current-voltage characteristics of SJJs in the deep BCS regime \cite{Anderson1969,Tinkham}. 
This enables an essentially model-free determination of the critical current $I_c$ at arbitrary coupling strengths, without requiring any \emph{a priori} knowledge of the system's many-body properties. %
A quantitative comparison between the measured $I_c$ throughout the crossover and the predictions of a newly developed analytic model allows for extracting the interaction-dependent order-parameter amplitude, finding good agreement with results from non-perturbative numerical many-body approaches \cite{Haussmann2007,Astrakharchik2005} to the BCS-BEC crossover problem. 
Our study highlights how transport measurements provide a powerful probe even for highly correlated quantum matter.

Our Josephson junction consists of two superfluid reservoirs comprising $N_{R,L} \simeq 3.5 \times 10^4$ atom pairs each, weakly coupled through a thin optical barrier (see Fig.~\ref{Fig1}A). The reservoirs are prepared by cooling a balanced mixture of the two lowest hyperfine states of ${}^6$Li below the condensation temperature at $T/T_F = 0.06(2)$, measured at unitarity \cite{note1}, where the superfluid critical temperature is $T_c/T_F \approx 0.21$ \cite{Haussmann2008}. The gas is initially confined into a cigar-shaped harmonic potential, with frequency ratios of about (1:14:12) along the $x$, $y$ and $z$-axis, respectively. Here, $T_F$ is the Fermi temperature given by $k_BT_F=E_F$, where $k_B$ is the Boltzmann and $E_F \simeq h \times 6$\,kHz is the Fermi energy of the non-interacting harmonically trapped gas, $h$ being the Planck constant. Interactions are parametrized by $(k_Fa)^{-1}$, where $a$ is the $s$-wave scattering length and $k_F=\sqrt{2m E_F}/\hbar$ is the Fermi wave vector, $m$ is the ${}^6$Li atomic mass and $\hbar=h/(2 \pi)$. 
We tune the scattering length between the two spin states via a broad Feshbach resonance located at 832\,G, accessing different superfluid regimes across the BCS-BEC crossover. The repulsive optical barrier at 532\,nm is shone along the $z$-axis and it has a Gaussian $1/e^2$ width $w\simeq 0.95(9)\,\mu$m along the $x$-direction, unless otherwise specified, while it is homogeneous along the $y$-direction. Its intensity profile and position are controlled by a digital micromirror device (DMD), whose surface is projected onto the atoms through a high $\mathrm{NA}\simeq0.5$ microscope objective. To initialize the junction at equilibrium, the barrier is initially located at the trap center $x_0$ and is raised to the target potential height $V_0$, experienced by one atom pair. This creates two identical reservoirs with relative population imbalance $z = (N_R-N_L)/N \simeq 0$, and correspondingly zero chemical potential difference, $\Delta \mu = \mu_{R} - \mu_{L}$. Here $N=N_R+N_L$ is the total pair number and  $\mu$ is the pair bulk chemical potential. Additionally, two DMD-generated repulsive light sheets select a $140\,\mu$m-long central region of the sample (see Fig.~\ref{Fig1}B).
These endcaps discard the most dilute, highest entropy regions of the sample, and aid to damp out any residual axial sloshing motion, conferring an excellent shot-to-shot stability below $1\%$ to the initially prepared number imbalance $z$.

We impose a pair current $I_{\mathrm{ext}}$ across the junction by setting the optical barrier in relative uniform motion with respect to the superfluid \cite{Giovanazzi2000}, as pictorially shown in Fig.~\ref{Fig1}A. With a constant total barrier displacement $\delta x\simeq10\,\mu$m, $I_{\mathrm{ext}}=\bar{z}\,N/2 \times |v|/\delta x$, where $v$ is the barrier velocity and $\bar{z} \simeq \mp0.15$ is the imbalance at equilibrium for the final barrier position $x_0 \pm\delta x$. To obtain the $I-\Delta\mu$ relation of the junction, we vary the $I_{\mathrm{ext}}$ and measure via in-situ imaging the imbalance $z$ at time $t_f=\delta x/|v|$, when the barrier translation is completed, obtaining thus $\Delta\mu=(z-\bar{z})\, E_c\,N/2$. Here, $E_{c} = 2\partial \mu_{L} / \partial N_{L}$ (calculated with $N_L = N/2$) is the effective charging energy of the junction, i.e., the inverse compressibility of the gas, and it reflects the density change between the two reservoirs due to the particle current through the barrier \cite{Meier2001,Giovanazzi2000}. For $|I_{\mathrm{ext}}|$ smaller than a critical value $I_c$, we observe pairs to tunnel coherently through the barrier [see panel (i) in Fig.~\ref{Fig1}B], maintaining $\Delta\mu = 0$ (i.e., $z-\bar{z}=0$). 
Conversely, for $|I_{\mathrm{ext}}| > I_c$ a finite $\Delta \mu$ develops, associated with the density increase arising from compression of the smaller reservoir [see panel (ii) in Fig.~\ref{Fig1}B]. 
Exploiting matter-wave interference between the expanding reservoirs, we can connect the observed pair tunnelling to the behavior of the phase difference $\varphi = \varphi_R - \varphi_L$. In Fig.~\ref{Fig1}C-D, we display typical complete measurements of both the $I-\Delta\mu$ and $I-\varphi$ characteristics of our junction for a molecular BEC (mBEC). Albeit with different current scale, we obtain the same $I-\Delta\mu$ relation near unitarity and in the BCS side of the Feshbach resonance. In particular, a non-linear response is clearly visible in the $I-\Delta\mu$ curve, with $\Delta\mu$ exhibiting a zero-resistance plateau for $|I_{\mathrm{ext}}| < I_c$ (see Fig.~\ref{Fig1}C). The value of $I_c$ is marked by the sharp onset of a chemical potential difference $\Delta\mu \neq 0$, after which the junction displays a resistive behavior. Correspondingly, $\varphi$ displays a non-linear monotonic increase, adjusting itself to sustain a supercurrent $|I_s| = |I_{\mathrm{ext}}| \leq I_c$ (see Fig.~\ref{Fig1}D). For sufficiently large $V_0$, as that used in Fig.~\ref{Fig1}C-D, $I_s (\varphi) \approx I_c \sin \varphi$, a signature not observed thus far in degenerate atomic gases \cite{Eckel2014b}. These observations unambiguously demonstrate that we access the Josephson dc regime, 
and that the tunnelling current observed below $I_c$ is a Josephson supercurrent. We have directly checked the departure from such sinusoidal $I-\varphi$ relation upon decreasing $V_0$ \cite{Bloch1970,Spuntarelli2007,Piazza2010}, observing the crossover to a linear current-phase characteristic peculiar of hydrodynamic weak links \cite{Hoskinson2006,Eckel2014b,Piazza2010,SM}. %
Moreover, we find that reversing the direction of the supercurrent across the junction causes a change of the sign of $\varphi$, i.e., $I_s(\varphi) \simeq -I_s(-\varphi)$, a feature directly associated to the order-parameter time-reversal symmetry expected in the present case of $s$-wave non-chiral superfluids.
To quantitatively describe the observed $I-\Delta\mu$ response, we employ the resistively and capacitively shunted junction (RCSJ) model widely applied for SJJs \cite{Stewart1968,McCumber1968,Tinkham,BaroneBook,SM}, namely a lump element circuit model incorporating a capacitive channel $C$ and a resistive channel $R$. The latter opens for $|I_{\mathrm{ext}}| > I_c$, not affecting the coherent dc Josephson branch, and incorporating any incoherent currents across the junction that induce a finite $\Delta\mu$. 
The capacitance $C=1/E_{c}$ is provided by self-consistent numerical calculations for crossover superfluids \cite{Haussmann2007,SM}. Within the RCSJ model, the dynamics of the junction is described by the Kirchhoff's law and the Josephson-Anderson relation:
\begin{align*}
&\frac{d^2\varphi}{d\tau^2}+\sqrt{\frac{1}{\beta_c}}\frac{d\varphi}{d\tau}+\sin{\varphi}=\frac{I_\mathrm{ext}}{I_c},\,
&\dot{\varphi} = - \Delta \mu/\hbar.
\end{align*}
Here, $\tau=\omega_p\,t$, where $\omega_p=\sqrt{I_c/(\hbar C)}$ is the plasma frequency. Such phase evolution is equivalent to the motion of a particle with mass proportional to $C$ and damping proportional to $G = R^{-1}$ in the tilted washboard potential $U(\varphi) = \hbar I_c\,(1 - \cos\varphi - (I_{\mathrm{ext}}/I_c)\,\varphi)$ \cite{BaroneBook,Tinkham}, and becomes classical for $\hbar \omega_p \ll \hbar I_c$ (phase regime) as in our case. The Stewart-McCumber parameter \cite{Stewart1968,McCumber1968} $\beta_c = I_c\,C/(\hbar G^2) \equiv Q^2$, where $Q$ is the quality factor of the junction, determines whether the oscillatory evolution $\varphi(t)$ is underdamped ($\beta_c \gg 1$) or overdamped ($\beta_c \ll 1$), yielding an hysteretic or non-hysteretic $I-\Delta\mu$ curve \cite{Tinkham}, respectively. The RCSJ model excellently captures the experimental $I-\Delta \mu$ characteristics as shown in Fig.~\ref{Fig1}C, enabling to extract the critical current $I_c$ and the conductance $G$, left as the only fitting parameters. The extracted $I_c$ values are negligibly sensitive to $E_c$, affording thus a direct and accurate determination of the critical currents with respect to previous studies \cite{Valtolina2015,Burchianti2018}. 

\begin{figure*}[t]
\centering
\includegraphics[width=120mm]{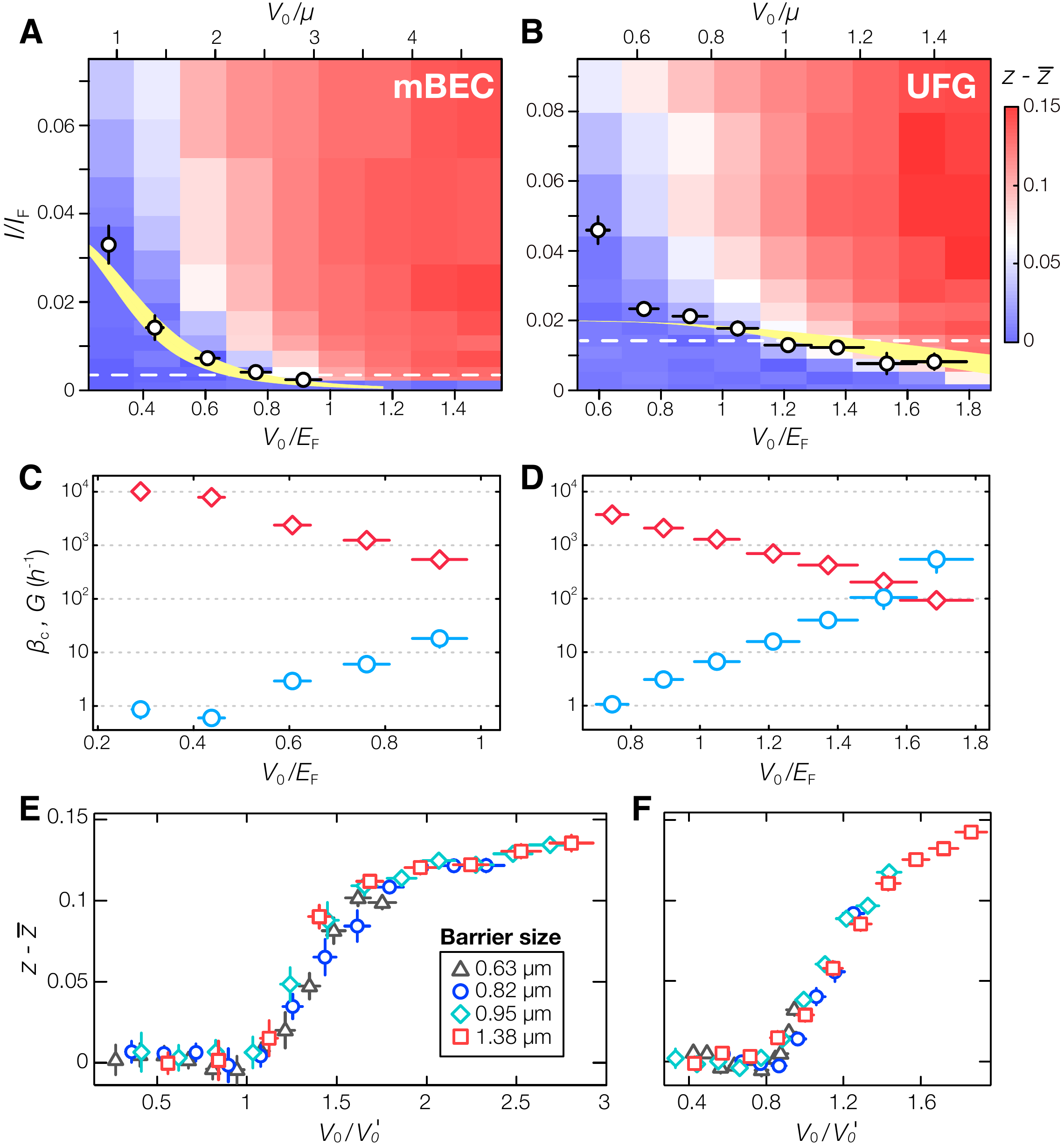}
 \caption{
 Dc Josephson effect in a tunable, ultracold Josephson junction.
 Current-imbalance characteristics for (\textbf{A}) a molecular BEC (mBEC), and (\textbf{B}) a unitary Fermi gas (UFG). Symbols denote the values of $I_c$ extracted through RCSJ-model fits, normalized to $I_F \equiv I_\mathrm{ext}({v=v_F})$, i.e., the current associated with a barrier moving at the Fermi velocity $v_F$. We note that the extracted values of $I_c$ always correspond to barrier motions much slower than the sound velocity, e.g.~$c\simeq0.35\,v_F$ at unitarity, as expected for high potential barriers. The vertical error bars denote the standard error of the fitting combined with uncertainty on $v_F$. The yellow shaded regions indicate the calculated $I_c$ (see text), considering a $10\%$ uncertainty around the nominal barrier width $w=0.95\,\mu$m. Condensate fractions $\lambda_0$ of 1 and 0.51 are assumed for the mBEC and the UFG, respectively. (\textbf{C}-\textbf{D}) Conductance $G$ (red diamonds) and Stewart-McCumber parameter $\beta_c$ (blue circles) as a function of $V_0/E_F$. (\textbf{E}-\textbf{F}) $V_0$-imbalance characteristics obtained for fixed $I_\mathrm{ext}$ [dashed horizontal lines in (\textbf{A}) and (\textbf{B})] in a mBEC and a UFG for different barrier widths, indicated in the legend. 
 Vertical error bars denote the standard error of the mean, while the horizontal ones combine the uncertainties in the calibration of $V_0$ and $E_F$.}
\label{Fig2}
\end{figure*}

In Fig.~\ref{Fig2}A-B, we display examples of the $I-\Delta\mu$ curves and the obtained $I_c$ as a function of $V_0$, for a mBEC superfluid at $(k_Fa)^{-1}\simeq 4.2$ and a unitary Fermi gas (UFG) at $(k_Fa)^{-1}\simeq 0$, respectively. As expected, the Josephson critical current is strongly suppressed with increasing $V_0$ in both regimes, since it is proportional to the tunnelling amplitude $|t|$ of pairs between the two coupled condensates \cite{Tinkham,Meier2001,Zaccanti2019}, that is essentially independent of the strength of the pairing and decreases exponentially with increasing $V_0 > \mu$. Figure~\ref{Fig2}C-D shows $\beta_c$ and $G$ as a function of $V_0$ for mBEC and unitary superfluids, which exhibit exponential behaviors as well. %
More specifically, we find a quadratic scaling $G \propto I_c^2$, in agreement with the prediction for weakly interacting BECs \cite{Meier2001}, where dissipative normal currents are associated with the emission of Bogoliubov sound modes or localized vortex-like excitations, as observed in previous experiments with crossover superfluids \cite{Burchianti2018}. Note that this behaviour differs starkly from the linear relation $G \propto I_c$ within the Ambegaokar-Baratoff formula \cite{Ambegaokar}, typically observed in SJJs, where the normal state conductance is simply a measure of the barrier transmission probability of single fermions, not necessarily associated with dissipative processes.
We also note the wide tunability of the Stewart-McCumber parameter, which increases to values as large as $\beta_c \sim 10^3$ for the highest $V_0$ explored in the UFG thanks to a steeper dependence $\beta_c \propto G^{-3/2}$ with respect to BCS superconductors. In this underdamped regime ($Q\gg1$), our junction is expected to be highly hysteretic, a promising condition for the observation of Shapiro resonances in the current-potential characteristic under an ac current drive \cite{BaroneBook,Tinkham}. 

\begin{figure*}[t]
\centering
\includegraphics[width=90mm]{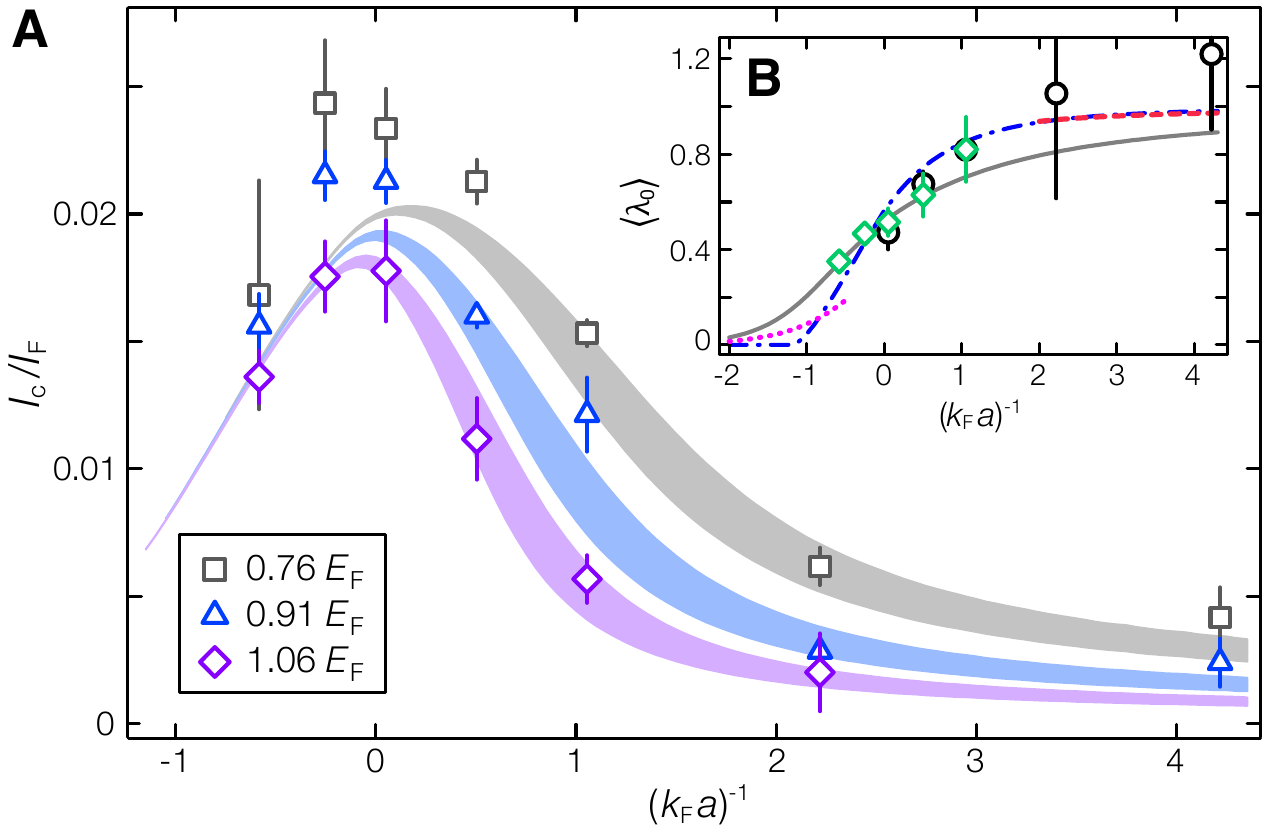}
 \caption{Josephson critical current and condensate fraction across the BCS-BEC crossover. 
 (\textbf{A}) Experimentally determined $I_c/I_F$ as a function of the interaction strength $(k_Fa)^{-1}$ for different barrier heights (symbols, see legend), together with the predictions of our analytic model (shaded areas), which employ the condensate fraction and chemical potential from Ref.~\citenum{Haussmann2007} and account for a $5\%$ uncertainty in the barrier width. Error bars are as in Fig.~\ref{Fig2}.
 (\textbf{B}) Total condensate fraction $\langle \lambda_0 \rangle$ obtained from $I_c$ including all data with $V_0/\mu > 0.6$ (black circles), or only the $V_0/E_F\simeq1.06$ data set (green diamonds). 
 $\langle \lambda_0 \rangle$ obtained by integrating homogeneous Luttinger-Ward \cite{Haussmann2007} and Quantum Monte Carlo results \cite{Astrakharchik2005} are plotted as a solid gray and dash-dotted blue line, respectively. 
 Dotted magenta and dashed red lines represent instead $\langle \lambda_0 \rangle$ calculated, respectively, from BCS theory including the Gorkov-Melik-Barkhudarov correction \cite{gor1961} and Bogoliubov quantum depletion for weakly interacting BECs \cite{RMPGiorgini}.}
\label{Fig3}
\end{figure*}

%

To gain a precise microscopic understanding of the observed behavior of the Josephson critical current, we rely on the analytic model recently presented in Ref.~\citenum{Zaccanti2019}. Within such a framework, expected to hold within the tunneling limit for any coupling strength throughout the BCS-BEC crossover, the critical pair current density per unit area can be expressed in terms of a bulk thermodynamic pre-factor and the single-pair barrier transmission amplitude. For a homogeneous junction with pair density $n$, this reads as
\begin{equation}
\hbar j_c=\frac{\mu\,n_c}{2k(\mu)}|t(\mu)|,
 \label{jchomo}
\end{equation}
where $k(\mu)=\sqrt{2M\mu}/\hbar$ is the wave vector of a bosonic pair of mass $M$ and energy $\mu$, $n_c = n \lambda_0$ is the density of condensed pairs, $\lambda_0$ is the condensate fraction, while $|t(\mu)|$ is the transmission amplitude of a single pair at incident energy $\mu$. 
Eq.~(\ref{jchomo}) can be extended to the harmonically trapped inhomogeneous case via the local density approximation (LDA) \cite{Zaccanti2019,SM}, obtaining predictions for the total $I_c$, as plotted in Fig.~\ref{Fig2}A-B.
To this purpose, we employ the local pair chemical potential $\mu(\mathbf{r})$ and condensate fraction $\lambda_0(\mathbf{r})$ based on non-perturbative Luttinger-Ward results for a zero-temperature homogeneous gas \cite{Haussmann2007}, while also accounting for second-harmonic contributions to the current-phase relation \cite{SM}. We find quantitative agreement with no free parameters, albeit expected discrepancies visible at small $V_0$ where hydrodynamic transport -- not included in Eq.~(\ref{jchomo}) -- becomes relevant. 
An additional validation of our model, and specifically of the separation between single- and many-particle properties in Eq.~\eqref{jchomo}, is presented in \mbox{Fig.~\ref{Fig2}E-F}. In order to isolate the barrier transmission contribution, we measure the imbalance $z-\bar{z}$ as a function of $V_0$ upon applying a fixed $I_{\mathrm{ext}}$ at various barrier widths $w$. By separately normalizing $V_0$ to the calculated barrier height $V_0'$ for which $I_c = |I_{\mathrm{ext}}|$ at each value of interaction and $w$, all different data sets collapse onto each other. This demonstrates that, for fixed interaction strength, the whole trend of $z-\bar{z}$ (and not only $I_c$) is predominantly determined by the tunnelling amplitude $|t|$.

We now turn to the measurement of $I_c$ at varying coupling strength $(k_Fa)^{-1}$, presented in Fig.~\ref{Fig3}A for three different fixed barrier heights $V_0/E_F$, allowing to disclose the order-parameter amplitude throughout the BCS-BEC crossover. Consistent with previous mean-field simulations \cite{Spuntarelli2007,Zou2014} and experimental studies of the Josephson plasma frequency \cite{Valtolina2015}, all data sets exhibit a non-monotonic trend of $I_c$ peaked around unitarity. We find our model to well reproduce $I_c$ for all interaction strengths, especially at the largest $V_0$. This agreement demonstrates how the condensate density $n_c$ (rather than the superfluid density), that quantifies pair long-range coherence \cite{Yang1962, Astrakharchik2005}, represents the key quantity setting the order-parameter amplitude reflected by $I_c$. Indeed, the monotonic decrease of $\mu$ with $(k_F a)^{-1}$ cannot explain the observed trend of $I_c$, and it is compensated by the increase of the condensate fraction $\lambda_0$, from exponentially small values in the BCS regime towards unity in the BEC limit \cite{Zwerger2012}.

A fairly accurate determination of the total condensate fraction $\langle \lambda_0 \rangle = N_c/N$, where $N_c = \int d^3r\,n_c(\mathbf{r})$ is the total number of condensed pairs in the trapped superfluid sample, can be obtained by approximating $I_c$ as \cite{SM}:
\begin{equation}
\hbar I_c \approx \langle \lambda_0 \rangle \times \int_{\mathcal{V}} d^3r\,n(\mathbf{r})\,\mu(\mathbf{r})\, \frac{|t(\mu(\mathbf{r}), V_0)|}{4 k(\mu(\mathbf{r})) R_x} \,,
 \label{Ic}
\end{equation}
where $R_x$ is the axial Thomas-Fermi radius of the cloud, $\mathcal{V}$ is the total junction volume, and $n(\mathbf{r})$ is the local density of pairs. %
We have numerically checked that the factorized Eq.~(\ref{Ic}), fully justified at unitarity within our temperature regime, %
approximates the $I_c$ calculated through LDA within a few percent for all interactions explored in our study \cite{SM}. %

From the measured $I_c$ at several coupling strengths and barrier heights,  
and by evaluating the integral in Eq.~(\ref{Ic}) \cite{SM}, %
we obtain $\langle \lambda_0 \rangle$ throughout the crossover region (see Fig.~\ref{Fig3}B). %
This is compared with non-perturbative $T=0$ predictions of $\langle \lambda_0 \rangle$, obtained by integrating $n_c(\mathbf{r})=\lambda_0(\mathbf{r})\,n(\mathbf{r})$ based on Luttinger-Ward \cite{Haussmann2007} (solid line) and Quantum Monte Carlo simulations \cite{Astrakharchik2005} (dot-dashed line). 
As a reference, we also plot the limiting trends of $\langle \lambda_0 \rangle$ calculated within the Bogoliubov and BCS approximation. 
We find our data to match the $T\!=\!0$ predictions quite well within the strongly interacting regime $|k_Fa|>1$.  %
In particular, we find $\langle \lambda_0 \rangle = 0.47(7)$ at unitarity, which agrees within its uncertainty with the value $\lambda_0=0.51$ from Ref.\,\citenum{Haussmann2007}, whereas it is significantly smaller than the mean-field result \cite{Salasnich2005} $\lambda_0 \simeq 0.7$, as well as the experimental value obtained at comparable temperatures [see e.g.~\cite{Horikoshi2010}] through the rapid-ramp technique \cite{Zwierlein2004,Regal2004}.

%
%

%
We experimentally demonstrated the intimate connection between dc Josephson currents and the complex order parameter %
in fermionic atom superfluids. %
Our work reveals that Josephson critical currents represent a reliable quantifier of the condensate density -- a fundamental microscopic quantity in any broken-symmetry superfluid state -- whose direct determination is typically hindered by strong interactions. %
We have checked the robustness of the critical current to temperature, observing no detectable decrease of $I_c$ at unitarity for $T\leq 0.1\,T_F$. An exciting extension will be to study the trend of $I_c$ and the nature of current-carrying excitations at $|I_\mathrm{ext}|>I_c$ upon approaching the superfluid critical temperature in three- and two-dimensional (possibly homogeneous) Fermi gases, contributing to a complete thermodynamic description of strongly interacting paired superfluids.
Furthermore, our experiments pave the way to exploring exotic current-phase relations in topological or tailored weak links, also in presence of external ac driving, and to unveiling low-temperature condensed phases in atomic simulators of the Fermi-Hubbard model.

\bigskip
\bigskip
\noindent\textbf{Acknowledgments}\\
We acknowledge inspiring discussions with E. Demler, L. Mathey, H. Moritz, A. Recati and J. Tempere. We thank S. Giorgini and G. E. Astrakharchik for providing us with QMC calculations of the condensate fraction across the BCS-BEC crossover, B. Frank for providing us with Luttinger-Ward calculations of the chemical potential and the condensate fraction across the BCS-BEC crossover, and H. Moritz and T. Lompe for careful reading of the manuscript. This work was supported under European Research Council GA no. $307032$ QuFerm2D, and no. $637738$ PoLiChroM, Fondazione Cassa di Risparmio di Firenze project QuSim2D 2016.0770, and European Union's Horizon 2020 research and innovation programme under the Marie Sk\l{}odowska-Curie GA no.~705269 (F.S.) and GA no. 843303 (W.J.K.). Correspondence and requests for materials should be addressed to F.S.~(E-mail: scazza@lens.unifi.it).\\

\vspace*{-10pt}

\bibliographystyle{Science} 

\makeatletter
\renewcommand\@biblabel[1]{#1.}
\makeatother
\bibliography{JosephsonBiblio}

\onecolumngrid

\newpage
\begin{center}
\textbf{
Supplementary Materials for\\[4mm]
\large Strongly correlated superfluid order parameters from dc Josephson supercurrents}\\
\vspace{4mm}
{W. J. Kwon$^{1,2}$, 
G. Del Pace$^{2,3}$, 
R. Panza$^{1,2}$, 
M. Inguscio$^{1,2,4}$, 
W. Zwerger$^{5}$,
M. Zaccanti$^{1,2}$, 
F. Scazza$^{1,2,\ast}$, 
and G. Roati$^{1,2}$}\\
\vspace{2mm}
{\em \small
$^1$Istituto Nazionale di Ottica del Consiglio Nazionale delle Ricerche (CNR-INO), 50019 Sesto Fiorentino, Italy\\
$^2$\mbox{European Laboratory for Nonlinear Spectroscopy (LENS), 50019 Sesto Fiorentino, Italy}\\
$^3$\mbox{Department of Physics and Astronomy, University of Florence, 50019 Sesto Fiorentino, Italy}\\
$^4$Department of Engineering, Campus Bio-Medico University of Rome, 00128 Rome, Italy\\
$^5$Technische Universit\"at M\"unchen, Physik Department, 85747 Garching, Germany\\[2mm]}
{\small$^\ast$ Corresponding author. E-mail: scazza@lens.unifi.it}\\
\end{center}

\bigskip

\setcounter{equation}{0}
\setcounter{figure}{0}
\setcounter{table}{0}
\setcounter{section}{0}
\makeatletter
\renewcommand{\theequation}{S.\arabic{equation}}
\renewcommand{\thefigure}{S\arabic{figure}}
\renewcommand{\thetable}{S\arabic{table}}
\renewcommand{\thesection}{S.\arabic{section}}

\setlength{\belowcaptionskip}{0pt}
\flushbottom
\linespread{1.25}

\section{Experimental methods}

\subsection{Sample preparation}
We prepare fermionic superfluids by evaporating a two-component mixture of the lowest hyperfine states of $^6$Li, confined within a crossed optical dipole trap. We employ the $\ket{F = 1/2, m_F = \pm 1/2}$ states, labelled as $\ket{1}$ and $\ket{2}$. Following the procedure reported in Refs.~\citenum{Burchianti2014,Valtolina2015,Burchianti2018}, the atomic sample is cooled at a magnetic field $B \simeq 832$\,G, on top of the $\ket{1}$-$\ket{2}$ Feshbach scattering resonance. In this way, we produce atomic samples of $N_t \simeq 1.3\times 10 ^5$ atoms per spin state at $T/T_F =0.06(2)$, measured using the known equation of state of a harmonically trapped unitary Fermi gas. At the end of the evaporation, we ramp the magnetic field adiabatically to the desired target value, which allows us to finely tune the inter-atomic $s$-wave scattering length $a$, whose magnetic-field dependence is taken from Ref.~\citenum{Zurn2013}. %
The overall harmonic potential is characterized by the trap frequencies $\omega_{x, y, z} \simeq (12, 165, 140)$ Hz. The magnetic contribution to the harmonic confinement depends on the magnitude of the Feshbach field, and a total variation of trap frequencies by $5 \%$ arises when spanning from the BEC regime at $B \simeq 702$\,G to the BCS regime at $B \simeq872$\,G.

\subsection{DMD projection and imaging setups}

\begin{figure*}[t]
\centering
\includegraphics[width=110mm]{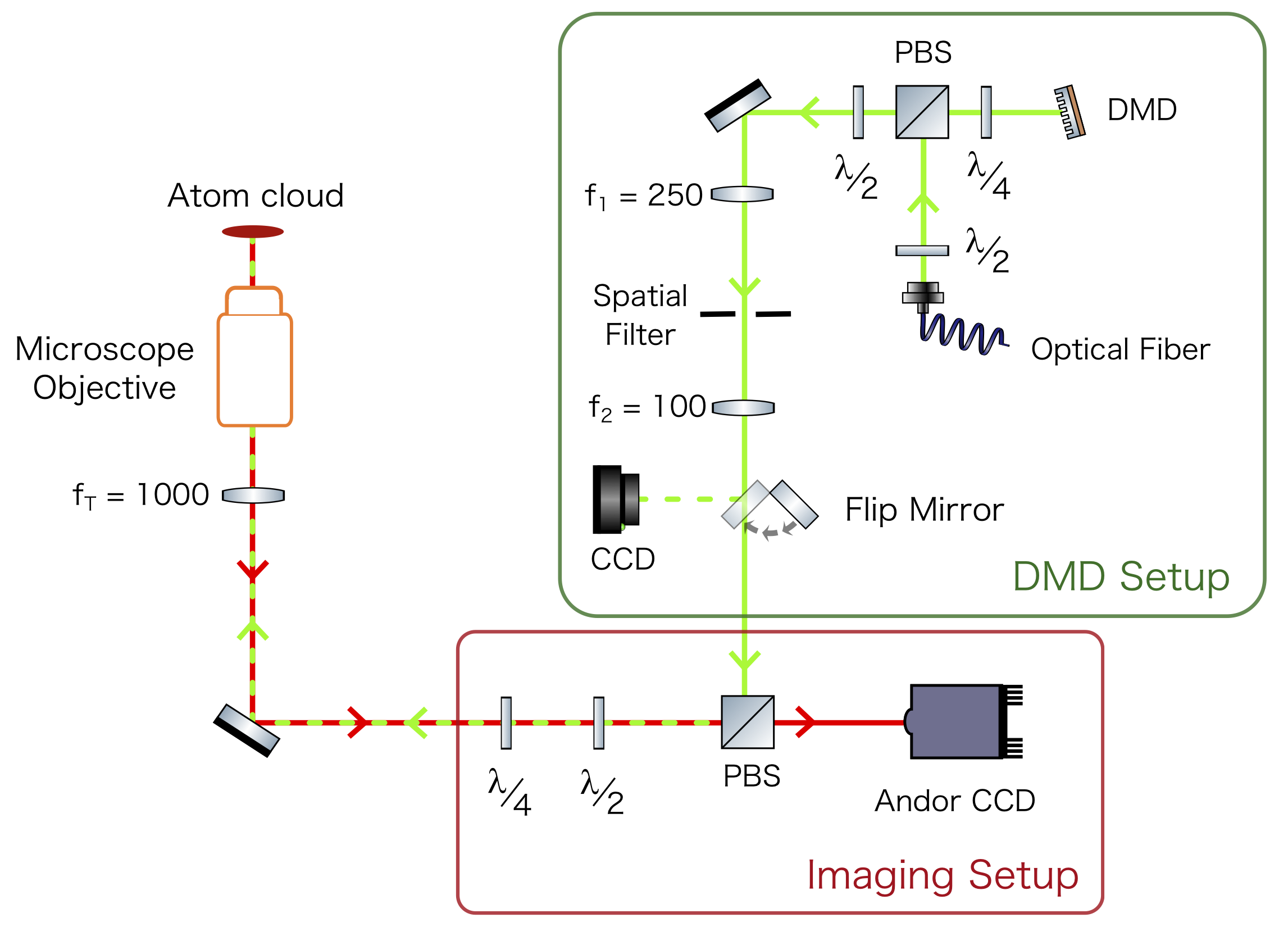}
 \caption{Sketch of the optical systems for high-resolution absorption imaging and DMD-generated arbitrary potential projection. The image of the atomic cloud is collected by a high resolution microscope objective (NA $\simeq 0.5$) and focused onto an Andor EMCCD camera by a $f_T = 1000$ mm tube lens. The DMD surface is imaged on the atomic cloud in two steps. It is first de-magnified by the $f_1 = 250$\,mm -- $f_2 = 100$\,mm telescope, and subsequently imaged onto the atomic plane through the tube lens and the objective, after being recombined with the imaging path by a PBS. Using a flippable mirror, we can record the DMD-created light pattern in the intermediate imaging plane with a Thorlabs CMOS camera, which allows for running the feedback algorithm and creating tailored optical potentials. In the Fourier plane of the first telescope, we use a tunable iris aperture as a spatial filter to reduce the numerical aperture of the DMD optical system independently from the imaging one.}
\label{DMDimagingSetup}
\end{figure*}

The tunneling barrier is created using a Digital Micromirror Device (DMD), equipped with the Texas Instruments Discovery 4100 0.7" XGA 2xLVDS (DLP7000) chip, provided by Vialux in the integrated V-7000 module. The chip is composed of a $1024 \times 768$ array of square micromirrors, with a pitch of $13.68\,\mu$m. The light pattern created by the DMD is imaged onto the atomic cloud along the vertical direction through a high-resolution imaging setup, as illustrated in Fig.~\ref{DMDimagingSetup}. This is based on a custom-made high-resolution microscope objective, featuring the same focal length for light at $670$\,nm and $532$\,nm. We use the objective both for imaging the density of the atomic cloud using resonant-light absorption imaging at $670$\,nm, and to imprint blue-detuned DMD-created optical potentials at $532$\,nm. 
The high numerical aperture of the microscope objective (NA $\simeq 0.5$) guarantees a sub-micrometer resolution of the imaging system at the used wavelengths, i.e., approximately $0.9\,\mu$m at $670$\,nm and $0.7\,\mu$m at $532$\,nm. 

\paragraph*{High-intensity absorption imaging}
All data presented in the main text have been collected using resonant absorption imaging along the vertical direction. %
Since the atomic cloud is optically dense, for in-situ absorption we use a high-intensity imaging beam with intensity $I/I_s \simeq 3$, where $I_s$ is the saturation intensity of the ${}^1S_{1/2} \rightarrow\, {}^3P_{3/2}$ imaging transition, also to keep the imaging pulse time sufficiently short to avoid unwanted effects ($4\,\mu$s pulse). We have calibrated our high-intensity imaging technique using the procedure described in Ref.~\citenum{Reinaudi2007}, while the magnification $21.82$ of the imaging optical system was calibrated by measuring the displacement of the cloud when finely tuning the waist position of one of the optical trap beams. %

\paragraph*{DMD projection setup}
The DMD is illuminated with a collimated, large-area Gaussian beam at $532$\,nm with a large waist of about $0.7$\,cm, exiting the output collimator of a large-mode optical fiber. The DMD surface is tilted by about $12$° with respect to the direction of the impinging illumination beam to make the reflected beam collinear with the incident one when all mirrors of the DMD are in the ON state. 
With this configuration we obtain a good illumination of the DMD surface and a high diffraction efficiency close to the blazing condition, while the small tilting angle prevents image distortions in the atomic plane. Incident and reflected beams are recombined through a polarizing beam-splitter (PBS), with the polarization set by $\lambda/2$ and $\lambda/4$ waveplates.

The DMD pattern is first de-magnified by the $f_1 = 250$ mm and $f_2 = 100$ mm telescope set in $f_1+f_2$ configuration, that provides a de-magnification of $2.52$. An adjustable iris is placed in the focal plane of the first lens, acting as a spatial filter to smoothen the discretized DMD image as required. The DMD projection path is subsequently combined with the imaging path on a large 2" PBS, and focused onto the atomic plane by the objective system. The DMD image is thus de-magnified by an overall factor $54.99$, with a single mirror of the DMD having a size of $0.25\,\mu$m in the atomic plane.

Directly after the first telescope, we have placed a flip mirror, allowing to focus the DMD image onto a Thorlabs CMOS camera for the purpose of calibrating the DMD pattern and creating arbitrary light profiles through a dedicated feedback algorithm. The feedback routine essentially operates with the same working principle of a PID feedback loop: it compares the image recorded by the camera with a target image, and minimizes the error between the two by applying a pixel-by-pixel error correction matrix on the DMD mirror array configuration. %
The mean number of ON mirrors in the DMD can be set in the feedback program, by setting the intensity of the target image. %
The quality of the feedback algorithm results is drastically improved by smoothing out as much as possible the intrinsic discreteness of the DMD image. This is done by setting the spatial filter, placed in between the telescope (see Fig. \ref{DMDimagingSetup}), to a small diameter of about a half millimeter while running the program. This corresponds to an effective low-pass spatial bandwidth of a couple of $\mu$m.

\vspace{-5pt}
\subsection{Barrier creation}

\begin{figure*}[t]
\centering
\includegraphics[width=80mm]{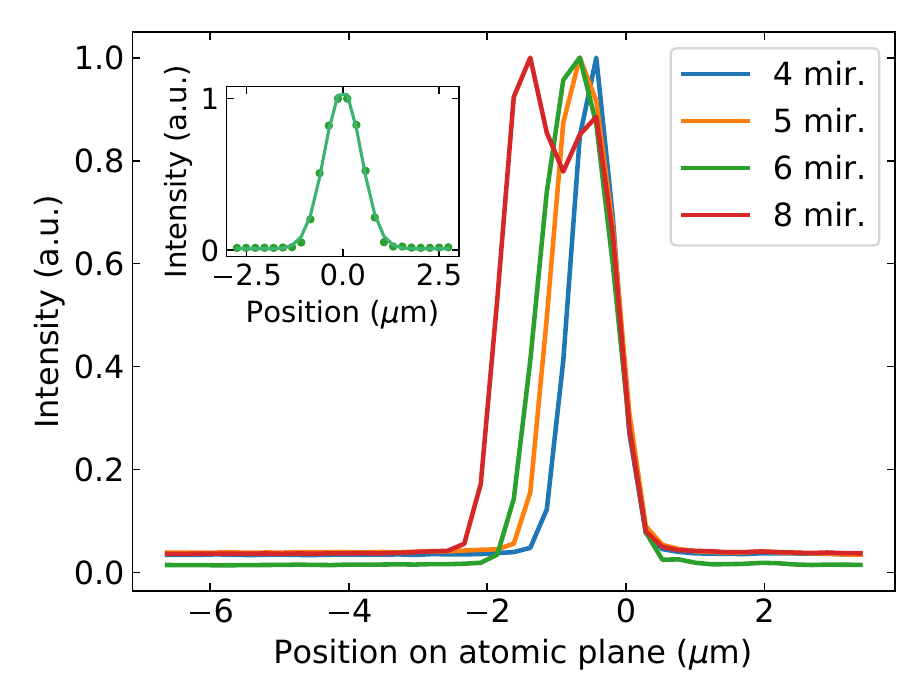}
 \caption{Comparison between barriers with different thickness, adjusted by varying the size of the DMD target image (expressed in terms of DMD mirror number -- see legend). From Gaussian fits of these profiles, we extract $1/e^2$ widths (radii) of $0.63$, $0.82$, $0.95$, and $1.38 \, \mu$m for the $4$, $5$, $6$ and $8$-mirror barriers, respectively. Inset: Gaussian fit of the $6$-mirror barrier axial profile used for measurements reported in the main text.}
\label{BarrierSize}
\end{figure*}

The tunneling barrier used for the measurements reported in the main text is created starting from a homogeneous light pattern, obtained through the above mentioned feedback routine. In particular, we run the feedback routine over a $300 \times 200$ mirrors region, setting the target output intensity to be constant at $45\%$ of the peak intensity of the Gaussian illumination beam. %
From the obtained target DMD image implementing the homogeneous pattern, we then cut out the rectangular region corresponding to the desired barrier potential.
The barrier width can be simply adjusted by varying the width of the rectangular cut-out region, which will be expressed in terms of number of DMD mirrors in the following.
Each barrier image is complemented by two lateral 100-mirror thick regions of all-ON mirrors, that are added along the axial direction. They provide the light sheets, or endcaps, that select the $140\,\mu$m axial central region of the atomic sample containing $N = N_R + N_L \simeq 7 \times 10^4$ atoms. By restricting the system to this region, we disregard the most dilute and highest entropy side regions of the cloud, while considerably improving the shot-to-shot stability. 
For all the experiments reported, the spatial filter is set to smoothen the barrier profile by cutting high spatial frequencies. This filter represents the limiting aperture of the DMD projection system, which we have confirmed by checking that the entire filtered DMD pattern is well collected within the tube lens and objective apertures. Therefore, we estimate the width of a created barrier by acquiring its image by the Thorlabs CMOS camera, which is a reliable approximation of the barrier image on the atomic plane. As reported in Fig.~\ref{BarrierSize}, we can vary the barrier $1/e^2$ width (radius) down to a minimum value of $0.63\,\mu$m for the $4$-mirror barrier, which is close to the limit set by the finite resolution of the optical system.

\begin{figure*}[b]
\centering
\includegraphics[width=80mm]{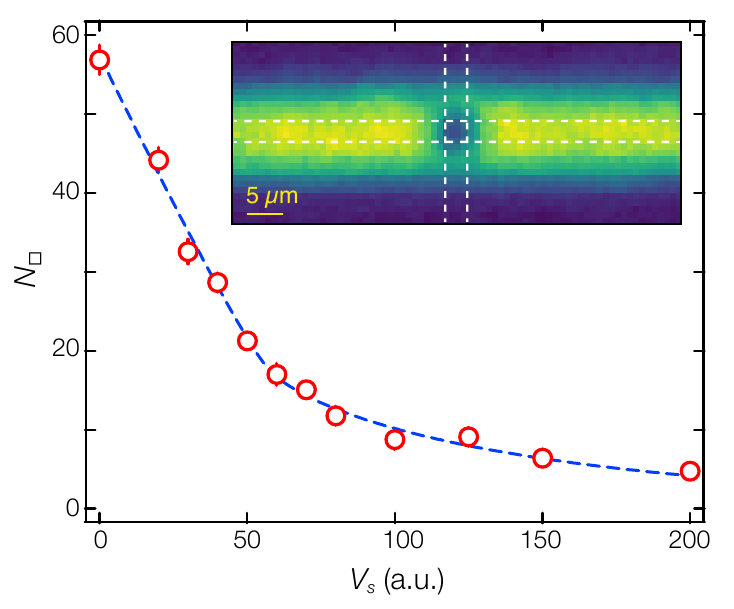}
 \caption{Calibration of the barrier height $V_0$ via the measurement of the density equation of state of a weakly interacting BEC. We apply a homogeneous square-shaped repulsive potential of $7.5 \times 7.5 \, \mu$m$^2$ in the center of a weakly interacting molecular BEC at $(k_F a)^{-1} \simeq 9$ (inset). We count the number of atoms $N_\square$ in the region delimited by the white dashed lines, for different powers of the DMD illumination beam. We fit the data (circles) with the semi-ideal model of a three-dimensional Bose gas (dashed line). We extract the power satisfying $V_s = \mu$, where $\mu$ is the pair chemical potential of the trapped gas evaluated using a weakly interacting BEC formula.}
\label{V0calib}
\end{figure*}

\subsection{Barrier height calibration}

\paragraph*{Barrier height calibration via density equation of state}
For the measurements reported in Fig.~2 and 3 of the main text, we vary the barrier potential height $V_0$ by controlling the beam power impinging on the DMD. The most accurate way to calibrate $V_0$ in our setup is to find out the power at which $V_0$ matches the accurately known chemical potential of a molecular BEC sample. The calibration method that we have implemented here is inspired by the measurement of the density equation of state $n(\mu)$ in two-dimensional homogeneous Fermi gases \cite{moritz2D}. We prepare a very weakly interacting molecular BEC at $(k_F a)^{-1}\simeq 9$, whose pair chemical potential $\mu$ can be precisely evaluated using the standard weakly interacting BEC formula. This has been also separately verified, by comparing the calculated $\mu$ with that extracted from the Thomas-Fermi radii obtained from bi-modal fitting of the in-situ density profile of a harmonically trapped sample.
Then, we apply a homogeneous square-shaped repulsive potential with an area of $7.5 \times 7.5 \mu$m$^2$ in the atomic plane, located at the center of the cloud (see the inset of Fig.~\ref{V0calib}). This potential pattern is produced through a cut-out of the same DMD image, corresponding to a homogeneous light pattern, also used for creating all barrier profiles. %
By measuring the change of local atom number $N_\square$ within the square region at varying height $V_s$ of the potential, we can extract the DMD illumination power for which $V_s = \mu$. For this, we fit the data with the %
semi-ideal density equation of state for a three-dimensional Bose gas \cite{dan}, as shown in Fig.~\ref{V0calib}. This calibration sets a well-defined reference for the barrier height $V_0$ felt by an atom pair, with an experimental uncertainty around 5\%, which is used for all experiments carried out at various $k_F a$ values. We repeated the same calibration procedure using a molecular BEC at $(k_F a)^{-1}\simeq 4.2$, obtaining a compatible result albeit with a larger uncertainty.

\paragraph*{Barrier height calibration via phase imprinting technique}

Here, we discuss an alternative method to calibrate the barrier height that confirms the result presented above. We use a phase imprinting technique to change the phase of the half of a molecular BEC ($(k_F a)^{-1} \simeq 4.2$), and obtain the calibration of $V_0$ by measuring the phase difference as a function of both time and power of imprinting pulse. %

\begin{figure*}[t]
\centering
\includegraphics[width=170mm]{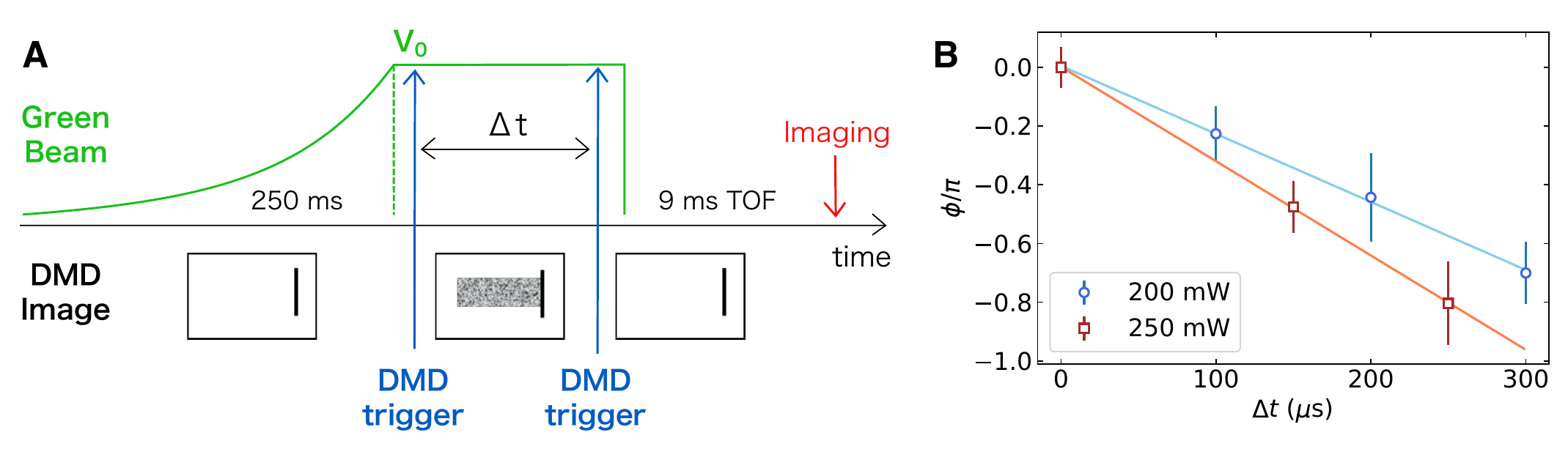}
 \caption{Sketch of the experimental procedure (\textbf{A}) and results (\textbf{B}) of the barrier height calibration via phase imprinting technique. (\textbf{A}) We adiabatically turn on the green light while displaying a barrier image on the DMD. When the barrier has reached the set value $V_0$, we switch the image to a composition of the barrier plus a $160 \, \mu $m$ \times 50 \, \mu$m homogeneous pattern of light, whose intensity is set to be a fixed fraction of the barrier intensity. After an imprinting time $\Delta t$, we change the DMD image back to the original barrier. We image the cloud after a $9$ ms time of flight and measure the imprinted phase from the emerging interference pattern. (\textbf{B}) Measured phase as a function of the imprinting time $\Delta t$ for different DMD illumation powers. A linear fit of the data yields the calibration of the optical imprinting potential $U$.}
\label{PIcalib}
\end{figure*}

A schematic diagram of the experimental sequence is reported in Fig. \ref{PIcalib} A. We adiabatically turn on a thin barrier of $\simeq 1 \, \mu$m width that bisects the cloud in two reservoirs, then we switch the image on the DMD (see next section for details) to a composition of the barrier plus a $160 \, \mu $m$  \times 50 \, \mu$m homogeneous pattern of light, whose height is set to be a fixed fraction of barrier height $V_0$. %
We keep the second image on for an imprinting time $\Delta t$, %
then we change the DMD image back to the barrier-only configuration. %
The acquired phase during $\Delta t$ is $\phi = U \times \Delta t/\hbar$, where $U$ is the light shift associated with the homogeneous pattern of light. 
We measure the imprinted phase $\phi$ by imaging the interference pattern arising from the two expanding clouds after a time of flight of $9$ ms \cite{Valtolina2015}. Specifically, $\phi$ is extracted by fitting the resulting interferogram by a 2D Gaussian modulated by a cosine function of the form $n(x,y)=A\exp{\{-x^2/w_x^2-y^2/w_y^2\}} \times \left(1+B \cos(kx+\phi)\right)$. As shown in Fig.~\ref{PIcalib} B, the imprinted phase linearly increases with the $\Delta t$ and with the power of the DMD illumination beam. From a linear fit, we obtain the light shift $U$ for a given power that is consistent with the calibration method described before.

\vspace{8pt}
\subsection{Barrier translation protocol}

\vspace{5pt}
To translate the barrier at a certain target velocity, we externally trigger the DMD to switch between subsequent images of barriers at different positions with the desired rate, the discreteness of the DMD setting the minimal displacement to $0.25\,\mu$m with our optical setup. To realize the total barrier displacement $\delta x \simeq 10\,\mu$m, we employ a sequence of 40 images. 
Each image consists of a barrier of fixed size displaced by $1$ DMD mirror from that in the previous image, superimposed to the all mirrors-ON regions to create the endcaps. 
By sending trigger signals with a constant delay time from one another, called the \textit{Picture Time}, the image displayed on the DMD is switched to the subsequent picture of the sequence, already uploaded on the board, with a time resolution of few $\mu$s. The barrier translation on the atomic cloud is thus a discrete movement whose velocity can be controlled by changing the \textit{Picture Time}. %

It should be noted that it is inevitable that a little light reaches the atomic cloud during the transition between two successive pictures, since all mirrors always first return to their rest position regardless of the subsequent image.  
Nonetheless, the total image switching time of the DMD is as small as $10\,\mu$s, and it is still considerably smaller than the minimum density response time of the sample $\sim \hbar/\mu \approx 30\,\mu$s, thus not affecting the cloud.

\begin{figure*}[t]
\centering
\includegraphics[width=170mm]{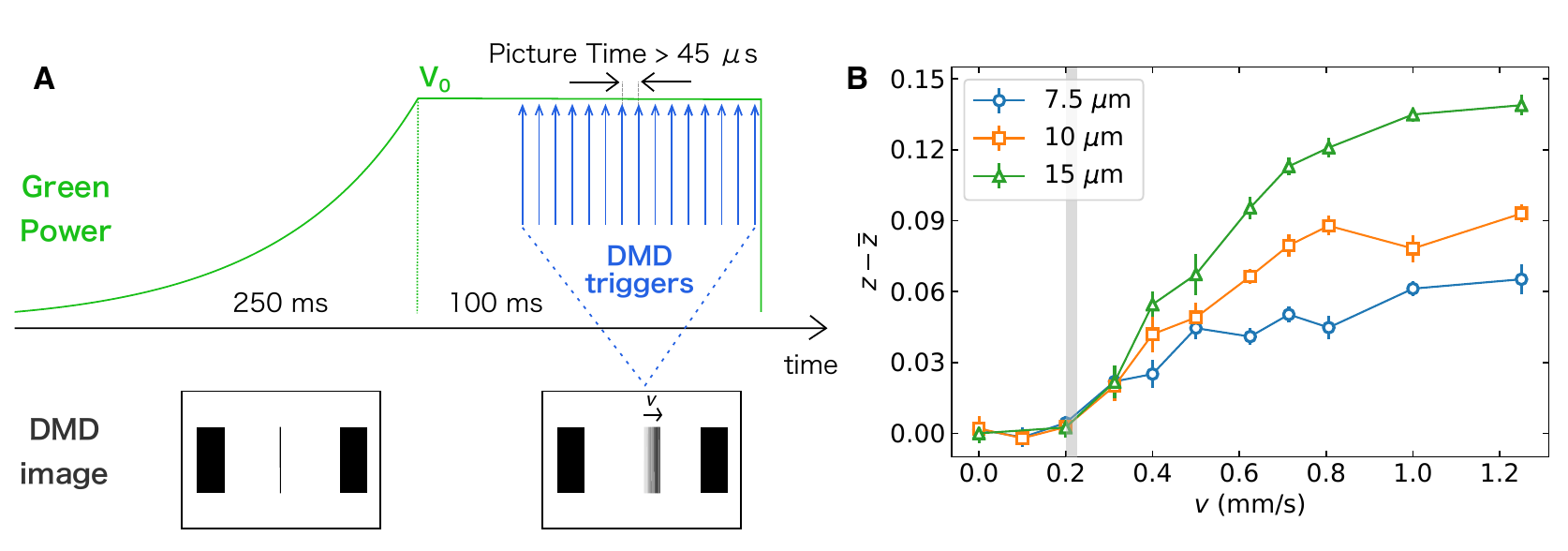}
 \caption{(\textbf{A}) Sketch of the experimental sequence to measure the critical current. In $250$ ms we adiabatically ramp up the green beam, while the DMD displaying the image of a centered barrier inside the $140 \, \mu$m region delimited by the two repulsive endcaps. When the barrier has reached the desired value $V_0$, we translate it by playing a sequence of images on the DMD. The switch between each picture and the following is controlled with a sequence of triggers equally spaced by the \textit{Picture time}, that sets the velocity of the barrier. %
 (\textbf{B}) Comparison of current-imbalance measurements for different total barrier translation length, performed in the mBEC regime ($(k_F a)^{-1} \simeq 4.2$) with a barrier waist $0.63 \, \mu$m and height $V_0/\mu \simeq 2.8$. We change the translation length by changing the total number of images in the played DMD sequence. All curves show a similar critical behaviour around $v \simeq 0.2$\,mm/s, as depicted by the shaded region.}
\label{DynProcedure}
\end{figure*}

The experimental procedure used for the measurements described in the main text is sketched in Fig. \ref{DynProcedure}A. We adiabatically ramp the DMD illumination power up over $250$\,ms to the desired value $V_0$, while displaying the first image of the sequence on the DMD. This picture corresponds to a barrier centered in the region delimited by the endcaps, and we finely tune the barrier position to the cloud center of mass to set the initial populations of the two reservoirs to be equal, i.e., $z=0$. 
When the green optical potential has reached the desired value $V_0$, we commence the barrier translation sending the desired sequence of triggers to the DMD. In particular, we use a sequence of $40$ equidistant triggers, %
yielding a movement of the barrier at constant velocity $v \simeq 0.25/t_P $ mm/s, where $t_P$ is the \textit{Picture Time} in ms. Since triggers are equidistant in time, the barrier velocity is sharply turned on to the desired value $v$.
We have also tried to smoothly increase the velocity up to the desired $v$ by adjusting the time spacing between the triggers to create an effective initial acceleration. We have verified that the two procedures do not give distinguishable results within our experimental resolution, so we employ the equidistant-trigger protocol to keep the translation time as short as possible, so as to render the contribution of incoherent pair transport across the barrier as small as possible. Whatever its origin, any incoherent current tends indeed to re-balance the reservoirs during the barrier translation itself, reducing the contrast of the measured $I-\Delta\mu$ curve. 

We further checked that the critical velocity (current) obtained from the imbalance measurement does not depend measurably on $\delta x$ for sufficiently small $\delta x$. 
In Fig.~\ref{DynProcedure}B, we show a comparison of the induced imbalance $z-\bar{z}$ after total barrier translations of $7.5$, $10$ and $15 \, \mu$m for a barrier $V_0/\mu \simeq 2.8$ in a BEC sample. %
We decided to use a total translation of $10 \, \mu$m, since it represents the best trade-off between a good signal to noise ratio and a short movement duration. %
Given the small barrier displacement with respect to the size of the reservoirs and the bulk properties of our reservoirs, these are expected to remain essentially in thermodynamic equilibrium at each point in time during the transport dynamics.

\section{Circuit model}

\begin{figure*}[t]
\centering
\includegraphics[width=75mm]{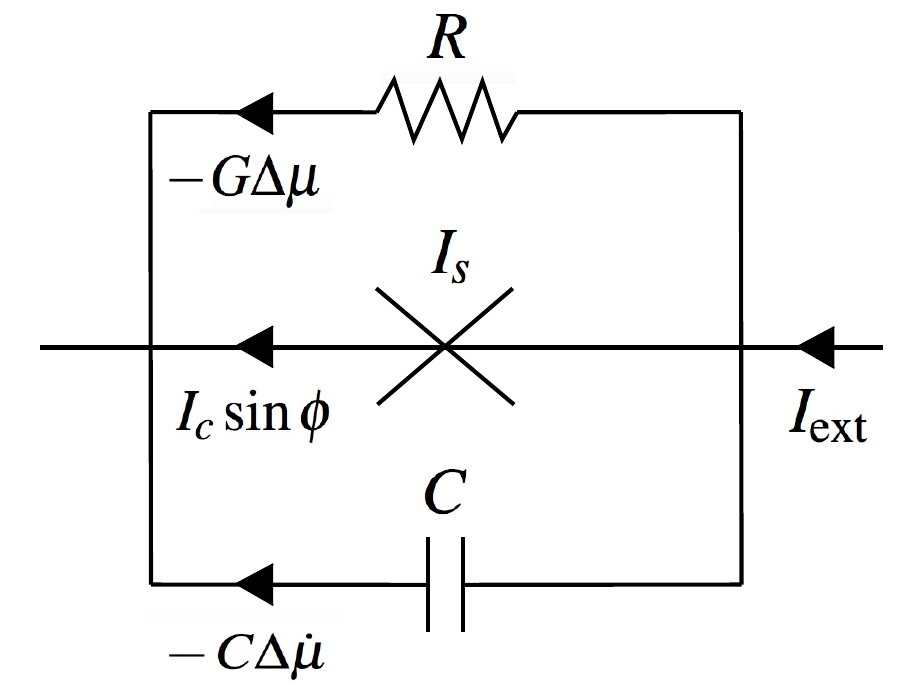}
 \caption{Sketch of the RCSJ circuit model used to quantitatively describe the $I- \Delta \mu $ response of the junction. The $\times$-shaped element represents the Josephson junction. The current flowing in each branch of the circuit is indicated close to the black arrows.}
\label{Circuit}
\end{figure*}

We characterize the $I - \Delta \mu$ response of the junction with the RCSJ circuit model, which is composed of three parallel elements injected by an external current $I_\mathrm{ext}$ (see Fig. \ref{Circuit}): a Josephson weak link with a current-phase relation $I_s = I_c \sin \varphi$, a shunt resistance $R$ and a capacitance $C$. The circuit is described by the two following coupled differential Josephson equations:
\begin{equation}
I_\mathrm{ext} = I_c \sin \varphi - G \Delta\mu - C\Delta \Dot{\mu},
\label{Eq:circuit_Iext}
\end{equation}
\begin{equation}
\hbar \Dot{\varphi} = - \Delta \mu,
\label{Eq:circuit_PhiDot}
\end{equation}
where $\varphi = \varphi_R - \varphi_L$ is the phase difference across the junction and $G = 1/R$ the conductance. The pair chemical potential difference $\Delta \mu = \mu_{R} -\mu_{L} $ is given by:
\begin{equation}
    \Delta \mu = \frac{N}{2} E_c \left( z - \bar{z} \right)
    \label{Eq:circuit_Dmu}
\end{equation}
where $N$ is the total number of atom pairs in the two reservoirs, %
$E_{c} = (\partial \mu_L/\partial N_L + \partial \mu_R/\partial N_R)$ is the charging energy of the junction \cite{Meier2001,Giovanazzi2000}, $z = (N_R-N_L)/N$ is the relative population imbalance and $\bar{z}$ is its equilibrium value. 
In our protocol we inject the current $I_\mathrm{ext}$ by displacing the barrier at constant velocity, so $\bar{z}$ changes over time during the translation depending on the direction of the barrier velocity $v$. In particular, given the small barrier displacement $\delta x$ used, which is below $7\%$ of the axial Thomas-Fermi radius of the cloud, and $\bar{z}(t=0) = 0$, we approximate the region swept by the barrier as homogeneous along the axial direction and write $\bar{z}(t) = \bar{z}_f\,t\, |v|/\delta x$, where $\bar{z}_f$ is the equilibrium imbalance when the barrier is in its final position at $t_f=\delta x/|v|$. Since we always discuss the imbalance resulting after the barrier translation is completed, we always refer for simplicity to $\bar{z}_f$ as $\bar{z}$ unless otherwise stated.

By combining Eqs. \ref{Eq:circuit_Iext} and \ref{Eq:circuit_Dmu}, %
 we obtain the following equation,
\begin{equation}
    \frac{N}{2} \Dot{z} = I_c \sin \varphi -G \Delta \mu.
    \label{Eq:circuit_zDot}
\end{equation}
We fit the experimental $I_\mathrm{ext}-\Delta \mu$ curves by numerically solving Eqs.~\eqref{Eq:circuit_PhiDot} and \eqref{Eq:circuit_zDot}, leaving only $I_c$ and $G$ as free parameters. $E_c$ is readily calculated within the local density approximation from self-consistent numerical results for crossover superfluids \cite{Haussmann2007}. It should also be noted that in the circuit model we assume a purely sinusoidal current-phase relation, regardless of typical (small) experimental deviations that do not affect the extraction of $I_c$. %

\begin{figure*}[h!]
\centering
\includegraphics[width=115mm]{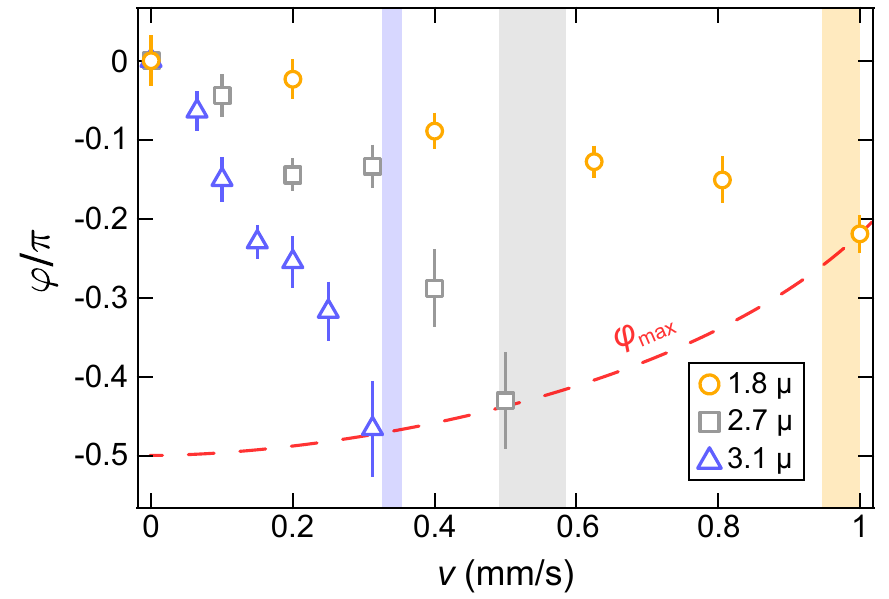}
    \caption{Current-phase relations measured using a barrier width $w\simeq0.63\,\mu$m and various barrier heights $V_0$. The phase is measured through matter-wave interference after a time of flight of $18$\,ms for a molecular BEC at $(k_F a)^{-1} \simeq 4.2$ (see Ref.~\citenum{Valtolina2015}). Shaded regions represent the standard confidence interval of the critical velocity, independently obtained by measuring the imbalance $z$ as a function of the barrier velocity $v$. The red dashed line is a guide to the eye for the value of the phase $\varphi_\mathrm{max}$ where the critical barrier velocity (current) is reached, qualitatively matching the expected trend \cite{Piazza2010}.}
\label{SM:CurrentPhase}
\end{figure*}

\section{Current-phase relation}

Neglecting spatial variations of the phase difference along the direction parallel to the junction boundary, $\varphi$ is a single number $\varphi \in [-\pi, \pi)$. The current-phase relation $I_s(\varphi)$ for a time-reversal invariant superfluid tunnel junction can always be expressed as a Fourier series $I_s(\varphi)=\sum_{n=1}^{\infty} I_n \sin(n \, \varphi)$, where the coefficients $I_n$ are linked to the tunneling amplitude for $n$ pairs to coherently traverse the barrier \cite{Bloch1970}. For sufficiently strong tunneling barriers featuring transmission probabilities $|t|^2 \ll 1$, only the $n=1$ term describing the tunneling of a single fermion pair remains non-negligible \cite{Bloch1970, Tinkham}, and Josephson's original result for an ideal junction holds, $I_s(\varphi)=I_c \sin(\varphi)$. For weaker barriers the $I_n$ decay instead slowly with $n$, leading to diverse current-phase relations. An extreme limit of such behavior, complementary to the tunneling limit, is the current-phase relation $I_s (\varphi) = I_c \sin (\varphi/2)$ (defined in the interval $-\pi < \varphi < \pi$ and periodically continued beyond) for a quantum point contact \cite{Beenakker1991}.

To explore the behaviour of the current-phase relation at various experimental barrier parameters in our geometry, we have measured $I_s(\varphi)$ with a reduced barrier Gaussian width of 0.63\,$\mu$m and varying barrier height $V_0$. The relative phase for each value of the applied current is extracted by fitting the interference pattern emerging from the two expanding reservoirs after a time of flight of $9$ ms (see insets of Fig.~1D in the main text), using a cosine-modulated 2D Gaussian profile, i.e.~$n(x,y)=A\exp{\{-x^2/w_x^2-y^2/w_y^2\}} \times \left(1+B \cos(kx+\varphi)\right)$. The thinner barrier used here facilitates the observation of the smooth change from a purely sinusoidal current-phase relation $I_c \sin\varphi$ to one with multiple harmonics, as the crossover is predicted to be slower with $V_0$ for thinner barriers \cite{Piazza2010}. We observe that for sufficiently high $V_0$ the current-phase relation remains close to ideal case. However, as we decrease the $V_0$, it deviates significantly from the sinusoidal form, approaching a linearly increasing trend for the weakest barrier \cite{Spuntarelli2007, Piazza2010}, where the maximum current occurs at $\varphi \approx \pi/5$ (see Fig.~\ref{SM:CurrentPhase}). Our phase measurement thus reveals that the current-phase relation indeed depends on the junction transmission properties, even for tunneling barrier heights well above the chemical potential, and that it is necessary to take into account higher harmonics to correctly describe the current-phase relation also for moderate barrier transmissions (e.g.~the lowest barrier in Fig.~\ref{SM:CurrentPhase}).

\section{Imbalance dynamics in the resistive (ac) branch}

\begin{figure}[t!]
\centering
\includegraphics[width=100mm]{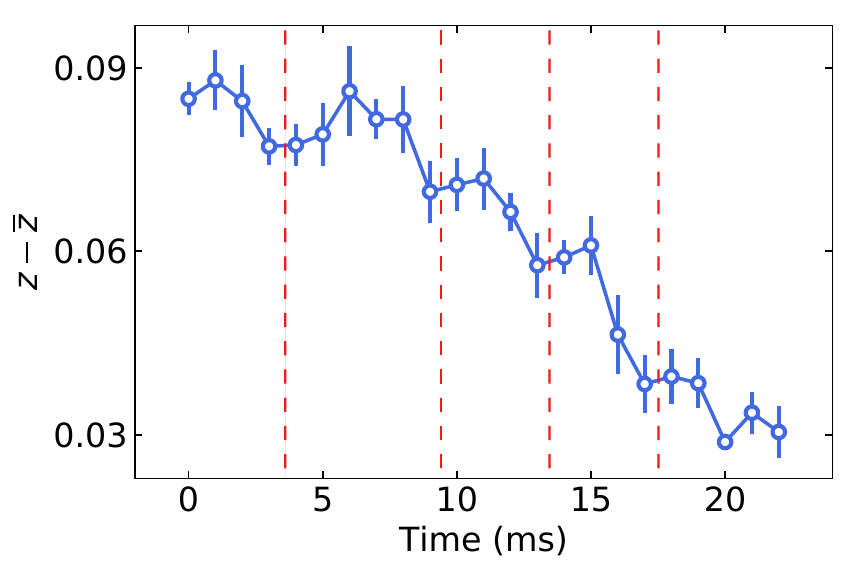}
 \caption{Post-translation dynamics of the imbalance in the resistive branch for a molecular BEC at $(k_F a)^{-1} \simeq 4.2$. The initial imbalance is created by moving a $0.63 \, \mu$m-wide barrier with $V_0/\mu \simeq 3.3$ at a velocity $v \simeq 1.0$\,mm/s above the critical one. The imbalance exhibits a decay towards $z = \bar{z}$ %
 (and subsequently oscillates around $\bar{z}$), consistently with Ref. \citenum{Burchianti2018}. However, one can distinguish a modulation on top of the decay of $z$ (red dashed lines mark visible minima in an otherwise monotonic decay), suggesting the occurrence of several phase slips \cite{Xhani2019}.}
\label{ACdynamic}
\end{figure}

When %
$I_\mathrm{ext}$ exceeds the critical current $I_c$, a chemical potential difference $\Delta \mu$ develops across the junction because of the finite barrier tunneling rate. In this section, we discuss the subsequent dynamics initiated by the non-zero $\Delta \mu$. 
Figure~\ref{ACdynamic} shows the imbalance time-evolution for a molecular BEC at $(k_F a)^{-1} \simeq 4.2$ after the translation of the $0.63 \, \mu$m barrier at $V_0/\mu \simeq 3.3$ with velocity $v = 1.0$ mm/s. Under these conditions $|I_{\mathrm{ext}}|>I_c$ and, at the end of the movement, i.e., at the beginning of the dynamics here investigated, $z - \bar{z} \neq 0$ (i.e.~$\Delta\mu\neq0$). In the absence of significant dissipation over the measurement time scale, one expects to observe ac Josephson-like oscillations \cite{Albiez2005, Levy2007}. Here, we instead observe the imbalance to decay down to a value $z -\bar{z} \simeq 0.02$, after which it undergoes small-amplitude oscillations around $\bar{z}$ (not shown). This dissipative behaviour is consistent with previous experimental measurements \cite{Burchianti2018}. As visible in Fig.~\ref{ACdynamic}, however, the decay is characterized by the presence of several minima. This is a remarkable experimental signature of phase-slippage processes (essentially connected to dissipative vortex nucleation \cite{Xhani2019}), that become visible thanks to the excellent shot-to-shot stability of the relative imbalance.

\section{Theoretical modeling methods}

In this section, we discuss the details of our microscopic theoretical description of the critical current $I_c$. 
By generalizing Eq.~(2) in the main text to the harmonically trapped, inhomogeneous case via the local density approximation (LDA), the total pair Josephson current $I_c$ can be written as (see Ref.~\citenum{Zaccanti2019} for more details in the derivation):
\begin{equation}
\hbar I_c = \int_{\mathcal{V}} d^3r \,n_c(\mathbf{r})\,\mu(\mathbf{r})\, \frac{|t(\mu(\mathbf{r}), V_0)|}{4k(\mu(\mathbf{r}))\,R_x} = \int_{\mathcal{V}} d^3r \,\lambda_{0}(\mathbf{r})\,n(\mathbf{r})\,\mu(\mathbf{r})\, \frac{|t(\mu(\mathbf{r}), V_0)|}{4k(\mu(\mathbf{r}))\,R_x},
 \label{realIc}
\end{equation}
where  $R_x$ is the axial Thomas-Fermi radius of the cloud, $\mathcal{V}$ is the total volume of the junction, $n_c(\mathbf{r})$ is the local condensate density of pairs, $\lambda_{0}(\mathbf{r})$ is the local condensate fraction, and $n(\mathbf{r})$ is the local pair density in the trap.  
$n(\mathbf{r})$ is assumed to coincide with the superfluid pair density, an approximation expected to hold at our low temperatures throughout the crossover region \cite{superfluidT,Sidorenkov2013}.
Only in the deep BEC limit $(k_Fa)^{-1} \gtrsim 4$, where the superfluid fraction coincides with the condensate one, and scales as $\lambda_0 \propto 1-(T / T_c)^3$, 
we employ the condensate density experimentally extracted through a bimodal fit to the in-situ density distribution, i.e., excluding a thermal component up to $\simeq 15\%$.

In order to evaluate Eq.~\eqref{realIc} in a unified manner across the BEC-BCS crossover, we exploit the polytropic approximation for the harmonically-trapped sample density profile, where the effective polytropic index $\gamma \equiv \partial \log\mu/\partial \log n$ is assumed to be constant, corresponding to the power-law relation $\mu \propto n^{\gamma}$. $\gamma$ is slowly varying with $(k_Fa)^{-1}$, and it is equal to 1 in the BEC limit, and to 2/3 for both unitary superfluids and in the BCS limit \cite{polytropic}. The polytropic approximation enables us to evaluate analytically $n(\mathbf{r})$ and $\mu(\mathbf{r})$ in the harmonic trap for any $(k_Fa)^{-1}$.

From the (bosonic) pair chemical potential of a homogeneous gas provided by Haussmann \textit{et al.} \cite{Haussmann2007}, $\mu_\mathrm{hom} = 2\epsilon_F\,\eta\!\left( (k_Fa)^{-1} \right)$ (where $\epsilon_F$ is the Fermi energy of a homogeneous system), we obtain $\gamma\equiv\gamma\left((k_Fa)^{-1}\right)$. The local chemical potential is then written as \cite{wenwen}
\begin{align}
 \begin{gathered}
 \mu(\mathbf{r})=\mu_{0}\left[1-\left(\frac{x}{R_x}\right)^2-\left(\frac{y}{R_y}\right)^2-\left(\frac{z}{R_z}\right)^2\,\right],\\
 \mu_{0}=2 E_F\left[\left[\eta\!\left((k_Fa)^{-1}\right) \right]^{1/\gamma}\times \frac{\sqrt{\pi}\,(1+\gamma)\, \Gamma(1/\gamma+5/2)} {8 \gamma\,\,\Gamma(1/\gamma + 2)}\right]^{2\gamma/(3\gamma+2)},
\label{mu}
\end{gathered}
\end{align}
with $\Gamma$ being the gamma function. The corresponding density profile is given by 
\begin{equation}
n(\mathbf{r}) =\frac{N}{2 \pi\,B(3/2,1/\gamma+1)\, R_x R_y R_z} \left[1-\left(\frac{x}{R_x}\right)^2-\left(\frac{y}{R_y}\right)^2-\left(\frac{z}{R_z}\right)^2\,\right]^{1/\gamma},
\label{density}
\end{equation} 
where $N$ is the total number of superfluid pairs in the harmonic trap, $B$ is the Euler beta function, and $R_{i}$ is the Thomas-Fermi radius along the $i$-direction. We remark that in order to evaluate Eq.~\eqref{realIc} for a given $(k_Fa)^{-1} \neq 0$ , the local Fermi vector $\kappa_{F}(\mathbf{r})=(6 \pi^2 n(\mathbf{r}))^{1/3}$ must be considered to properly take into account the 
spatial variation of $\lambda_{0} (\mathbf{r})$, reflecting the dependence of the local (homogeneous) condensate fraction upon $(\kappa_{F}a)^{-1}$.

\paragraph*{Second-harmonic current contribution} To estimate the critical current more accurately, we consider the second-harmonic contribution to the current-phase relation, namely $I_2 \sin{2\varphi}$. This correction is not negligible, especially when $V_0 < \mu$, %
and it increases the maximum Josephson supercurrent above the first-order $I_c$ given by Eq.~\eqref{realIc}. 
Following Ref.~\citenum{Meier2001}, the magnitude of $I_2$ in our trapped configuration can be calculated by replacing $|t|/4$ with $|t|^{2}/16$ in Eq.~\eqref{realIc}, using the same integral form:
\begin{equation}
\hbar |I_2| = \int_{\mathcal{V}} d^3r \,\lambda_{0}(\mathbf{r})\,n(\mathbf{r})\,\mu(\mathbf{r})\, \frac{|t(\mu(\mathbf{r}), V_0)|^{2}}{16\,k(\mu(\mathbf{r}))\,R_x} \,
 \label{2nd}.
\end{equation} 
Following Ref.~\citenum{goldobin}, the maximum Josephson current $I_\mathrm{max}$ can then be evaluated analytically as %
\begin{equation}
I_\mathrm{max} = f(g)\,I_c = \frac{(\sqrt{1+32g^2}+3)^{3/2} \,(\sqrt{1+32g^2}-1)^{1/2}}{32|g|}\,I_c,
 \label{Imax}
\end{equation}
with the correction factor $f(g)\geq 1$, where $g\equiv |I_2|/I_c$. Although the leading term $I_c$ given by Eq.~\eqref{realIc} represents the dominant contribution for the maximum supercurrent, we observe that $I_2$ can increase $I_\mathrm{max}$ by up to $10\%$ for the lowest barrier heights used in unitary and BCS regimes, i.e., $f\approx1.1$. Therefore, we take into account the contribution of the second-harmonic Josephson current in all theory/experiment comparisons presented in the main text, evaluating the maximum supercurrent through Eq.~\eqref{Imax}. 

\paragraph*{Barrier depth of focus} Since the depth of focus associated with the small width $w_0$ of the optical barrier is comparable to the thickness of our samples, we account for it in the evaluation of Eqs.~\eqref{realIc} and \eqref{2nd}. 
Specifically, this is done by approximating the barrier propagation along the $z$-axis with that of a Gaussian mode, focused at the center of the cloud ($z=0$) where waist (height) is set to $w_0$ ($V_0$). For $z\neq0$ the barrier width varies as $w \approx w_0 \sqrt{1+(z/z_R)^2}$ and the height as $V(z) \approx V_0/\sqrt{1+ (z/z_R)^2}$, where $z_R=\frac{\pi w_{0}^2}{\lambda}$ is the Rayleigh length associated to $w_{0}$.

\subsection{Total condensate fraction of trapped samples}
At low temperatures, the condensate fraction is a locally slowly varying quantity in harmonically trapped superfluids. %
Our measurement of the maximum Josephson current allows to obtain a good estimate of the total condensate fraction $\langle \lambda_0 \rangle$ in the inhomogeneous samples, defined as
$$\langle \lambda_0 \rangle \equiv {N_c \over N} = \frac{\int_{\mathcal{V}} d^3r \,\lambda_{0}(\mathbf{r})\,n(\mathbf{r})}{\int_{\mathcal{V}} d^3r \,n(\mathbf{r})}.$$

This is possible owing to the fact that the integrals of Eq.~\eqref{realIc} and \eqref{2nd} can be approximately factorized as 
\begin{equation}
I_c\simeq\langle \lambda_0 \rangle \times I_{c,\,\mathrm{sup}}\, \quad\text{and}\quad 
I_{2}\simeq\langle \lambda_0 \rangle \times I_{2,\,\mathrm{sup}},
\end{equation}
leading to a discrepancy of up to $3\%$ from the exact evaluation of Eq.~\eqref{realIc} and \eqref{2nd} across the experimentally explored range of interaction strengths, and thus justifying Eq.~(3) in the main text. 
Here, $I_{c,\,\mathrm{sup}}$ and $I_{2,\,\mathrm{sup}}$ are defined by setting $\lambda_0 (\mathbf{r}) \equiv 1$ in Eq.~\eqref{realIc} and Eq.~\eqref{2nd}, respectively. %
The relatively good accuracy of such approximation stems from the slowly varying behaviour of $\lambda_{0}(\mathbf{r})$ across the majority of the in-trap density distribution. 
Therefore, the theoretical prediction of the maximum Josephson current $I_\mathrm{max}$ can be suitably recast as 
\begin{equation}
I_\mathrm{max}=f\left(\frac{I_2}{I_c}\right) I_c \approx \langle \lambda_0 \rangle\, f\left(\frac{I_{2,\,\mathrm{sup}}}{I_{c,\,\mathrm{sup}}}\right)\, I_{c,\,\mathrm{sup}}.
\end{equation}
By employing this latter relation and the experimentally determined maximum Josephson current $I_{c,\,\mathrm{exp}}$, we extract $\langle \lambda_0 \rangle$ for each coupling strength across the BCS-BEC crossover as
\begin{equation}
\langle \lambda_0 \rangle\approx \frac{I_{c,\,\mathrm{exp}}}{f\left(\frac{I_{2,\,\mathrm{sup}}}{I_{c,\,\mathrm{sup}}}\right)I_{c,\,\mathrm{sup}}},
\label{avgcond}
\end{equation} 
Such extraction of the total condensate fraction $\langle \lambda_0 \rangle $ is reported for various coupling strengths in Fig. 3B of the main text. 

\subsection{Eckart barrier as an approximation of Gaussian barrier}
In order to evaluate the transmission amplitude $|t|$ of bosonic pairs through our Gaussian barriers, we approximate their longitudinal $x$-axis profile with a symmetric Eckart potential $V(x) = 1/\cosh^2{(x/d)}$, where $d=0.6\, w_0$. This enables us to analytically calculate $|t|$ as a function of the incident energy. %

\begin{figure*}[h]
\centering
\includegraphics[width=100mm]{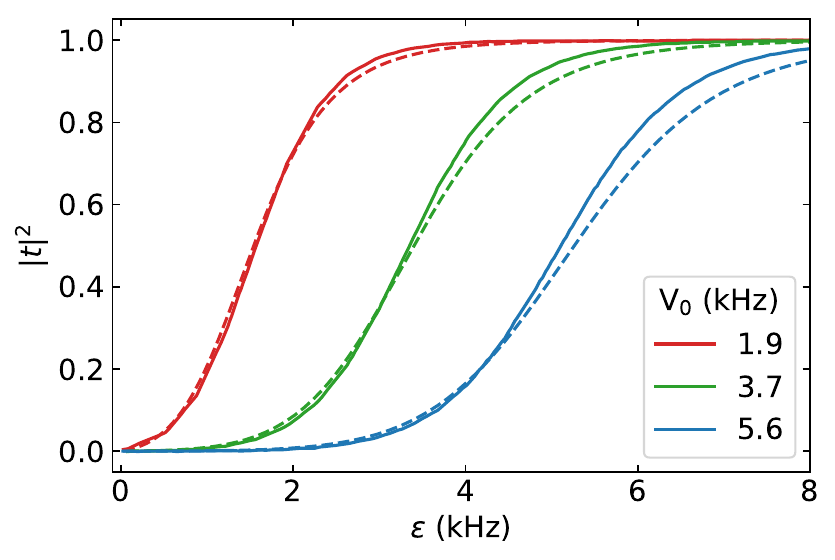}
 \caption{Comparison of the transmission probability through a Gaussian barrier of width $w = 0.95 \, \mu$m (solid lines) and an Eckart one with dimension $d = 0.6 w$ (dashed lines), for different barrier height $V_0$ as a function of the incident energy $\varepsilon$. Gaussian barrier transmission is obtained following the procedure described in Ref.~\citenum{Fernandez2011}, while the Eckart one is computed from the analytical formula in Ref.~\citenum{haar}.}
\label{SM:EckartBarrier}
\end{figure*}

To check the validity of such approximation, we compare the transmission probability $|t|^2$ of a single particle at energy $\varepsilon$ through Gaussian and Eckart barriers with fixed width $w$ and various potential heights, as displayed in Fig. \ref{SM:EckartBarrier}. We computed the transmission through the Gaussian barrier following the procedure described in Ref.~\citenum{Fernandez2011}, while we calculate the Eckart transmission from the analytic formula in Ref.~\citenum{haar}. In the regime of low energy $\varepsilon \lesssim 0.5\,V_0$, the two curves overlap reasonably well, while for high energies $\varepsilon > 0.5\,V_0$ the Eckart barrier transmission deviates from that of the Gaussian one. Nonetheless, the behavior of the Eckart barrier transmission follows that of the Gaussian barrier within a few percent relative error, which should not seriously affect any analysis presented in the main text. Moreover, in the experiment the barrier is typically set to yield pair transmission probabilities $|t|^2 < 0.5$, thus the energy range explored is that where the best agreement with the Eckart approximation is obtained.

\end{document}